\documentclass[a4paper,12pt]{article}
\usepackage[pctex32]{graphics}

\begin{document}
\topmargin 0pt \oddsidemargin 0mm

\renewcommand{\thefootnote}{\fnsymbol{footnote}}
\begin{titlepage}
\vspace{5mm}
\begin{center}
{\Large \bf New  agegraphic dark energy model with generalized
uncertainty principle} \vspace{12mm}

{\large   Yong-Wan Kim, Hyung Won Lee and Yun Soo
Myung\footnote{e-mail
 address: ysmyung@inje.ac.kr}}

\vspace{10mm} {\em  Institute of Basic Science and School of
Computer Aided Science \\ Inje University, Gimhae 621-749, Korea }
\\
 {\large and \\ Mu-In Park}
 \\
\vspace{10mm} {\em  Research Institute of Physics and Chemistry
\\Chonbuk National University, Chonju 561-756, Korea }
\end{center}

\vspace{5mm} \centerline{{\bf{Abstract}}}
 \vspace{5mm}

We investigate the new agegraphic dark energy models with
generalized uncertainty principle (GUP).  It turns out that
although the GUP affects the early universe, it does not change
the current and future dark energy-dominated universe
significantly. Furthermore, this model could describe the
matter-dominated universe in the  past only when the parameter $n$
is chosen to be $n>n_c$, where the critical value determined to be
$n_c=2.799531478$.

\end{titlepage}
\newpage
\renewcommand{\thefootnote}{\arabic{footnote}}
\setcounter{footnote}{0} \setcounter{page}{2}

\section{Introduction}
Observations of supernova type Ia suggest that our universe is
accelerating~\cite{Riess98,Perlmutter99,Riess04,Astier04}.
Considering the ${\rm \Lambda}$CDM
model~\cite{SDSS1,SDSS2,SDSS3,Wmap1,Wmap2,Wmap3}, the dark energy
and cold dark matter contribute $\Omega^{\rm ob}_{\rm
\Lambda}\simeq 0.74$ and $\Omega^{\rm ob}_{\rm CDM}\simeq 0.22$ to
the critical density of the present universe. Recently, the
combination of WMAP3 and Supernova Legacy Survey data shows a
significant constraint on the equation of state (EOS) for the dark
energy, $w_{\rm ob}=-0.97^{+0.07}_{-0.09}$ in a flat
universe~\cite{WMAP4,SSM}.

Although there exist a number of dark energy models~\cite{CST},
the two promising candidates are the cosmological constant  and
the quintessence scenario~\cite{UIS}. In order to resolve the
cosmological constant problem, we may need to introduce a
dynamical, cosmological constant model. The EOS for the latter is
determined dynamically by the scalar or tachyon.

Also there exist dynamical models of the dark energy which satisfy
the holographic principle. One is the holographic dark energy
model~\cite{LI} and the other is the agegraphic dark energy
model~\cite{CAI}. The first is based on the Bekenstein-Hawking
energy bound $E_{\rm \Lambda} \le E_{\rm BH}$ with the energy
$E_{\rm BH}$ of a universe-sized black hole which produces $ L^3
\rho_{\rm \Lambda}\le m_{\rm p}^2L$~\cite{CKN,myung} with the
length scale $L$ (IR cutoff) of the universe and the Planck mass
$m_{\rm p}$.  The latter is based on the K\'{a}rolyh\'{a}zy
relation of $\delta t$~\cite{Karo,ND,Sas,Maz} and the time-energy
uncertainty of $\Delta E\sim t^{-1}$ in the Minkowiski spacetime
with a given time scale $t$, which  gives  $\rho_{\rm q}\sim
\frac{\Delta E}{(\delta t)^3}\sim \frac{m_{\rm p}^2}{t^2}$. We
note that this expression of energy density  first appeared in
Ref.\cite{Sas}.
 Hence
we find the vacuum energy density $\rho_{\rm \Lambda}=3c^2m^2_{\rm
p}/L^2$ as the holographic dark energy density~\cite{HSU,HM},
whereas the energy density of metric perturbations $\rho_{\rm
q}=3n^2m^2_{\rm p}/T^2$ with the age of the universe $T=\int^t_0
dt'$ as the agegraphic dark energy density. Here the undetermined
parameters $c$ and $n$ are introduced to describe the appropriate
dark energy model.
 It seems that the agegraphic  dark energy model does not
suffer the causality problem of the holographic dark energy model
because the agegraphic  dark energy model do not use the future
event horizon~\cite{WC1,WC2,WZLCZ,ZLWWC,Neupane}. However, this
model suffers from the contradiction to describe the
matter-dominated universe in the far past. Hence, the new
agegraphic dark energy model with the conformal time
$\eta=\int^t_0 dt'/a(t')$ with the scale factor $a'=a(t')$ was
introduced to resolve this issue~\cite{WC3,WC4,Neupane07}.

 Nowadays we are interested in the generalized
uncertainty principle (GUP) and its
consequences~\cite{Garry,Scar,Rama,CMOT,Li02,Setare,MV,Nozari,HH06,LHL,MKP,MIPARK,KSY}
since the Heisenberg uncertainty principle is not expected to be
satisfied when quantum gravitational effects become important.
Even though the GUP has its origins in the string
theory~\cite{GM87,ACV,KPP90}, the GUP provides the minimal length
scale, the Planck scale $l_{\rm p}$ and may play a role of
evolution of the universe. Especially, we expect that this may
modify the evolution of early universe at the Planck scale and
inflation significantly.

In this Letter, we investigate  the new agegraphic dark energy
models with the GUP. We compare this  with new  agegraphic dark
energy models. Especially, we show that the parameter $n$ of  the
new agegraphic dark energy model with the GUP is restricted to
$n>n_c$ for $q=1$, in order to describe the matter-dominated
universe in the far past. As far as we know, this is the first
time to incorporate the GUP into the cosmology to explain the dark
energy-dominated universe.

\section{New agegraphic dark energy model with GUP}
We start with extending the time-energy uncertainty  to the
GUP~\cite{Garry,Scar,Rama,CMOT,Li02,Setare,MV,Nozari,HH06,LHL,MKP,MIPARK,KSY}
\begin{equation}
\Delta E \Delta t \ge 1+ \alpha (\Delta E)^2
\end{equation}
in the units of $c=\hbar=k_B=1$. Here the parameter $\alpha$ has
the Planck length scale like $\alpha \sim l^2_{\rm p}\sim
1/m^2_{\rm p}$. Solving the saturation of the GUP leads to
\begin{equation}
\Delta E_{\rm G} =\frac{1}{\Delta t}+\frac{\alpha}{(\Delta
t)^3}=\frac{1}{t}+\frac{\alpha}{t^3},
\end{equation}
where we use the relation of  $\Delta t \sim t$ for cosmological
purpose. Then the energy density inspired by the GUP is defined
by
\begin{equation}
\rho_{\rm G}=\frac{\Delta E_{\rm G}}{(\delta t)^3}
\end{equation}
where $\delta t$ is given by the K\'{a}rolyh\'{a}zy relation of
time fluctuations as $\delta t= t^{2/3}_{\rm
p}t^{1/3}$~\cite{Karo,Maz}. For the labelling of $\alpha$ and
$t_{\rm p}$ as
\begin{equation}
\alpha=\Big(\frac{q}{n}\Big)^2\frac{1}{m^2_{\rm p}},~t^2_{\rm
p}=\frac{1}{3n^2m^2_{\rm p}},
\end{equation}
respectively, the energy density  is  described  with two
parameters ($n,q$) as
\begin{equation}
\rho_{\rm G}=\frac{3n^2 m^2_{\rm p}}{t^2}+ \frac{3 q^2}{t^4}.
\end{equation}
The new  agegraphic dark energy model is described  by using the
conformal time $\eta$ instead of the age of universe $T$
\begin{equation}
\eta=\int^a_0\frac{da'}{(a')^2H'}=\int^x_{-\infty}\frac{dx'}{a'H'}
\end{equation}
with $x=\ln a$ and the Hubble parameter $H'=\dot{a'}/a'$. The
corresponding energy density takes the form
\begin{equation}
\rho_{\rm G}=\frac{3n^2 m^2_{\rm p}}{\eta^2}+ \frac{3
q^2}{\eta^4}.
\end{equation}

A flat universe composed of $\rho_{\rm G}$ and the cold dark
matter $\rho_{\rm m}$ is governed by the first Friedmann equation
\begin{equation} \label{fried}
H^2=\frac{1}{3m^2_{\rm p}}(\rho_{\rm G}+\rho_{\rm m})
\end{equation} and their continuity equations
\begin{eqnarray} \label{cont1}
\dot{\rho}_{\rm G}+3H(\rho_{\rm G}+p_{\rm G})&=&0,\\
\label{cont2}\dot{\rho}_{\rm m}+3H\rho_{\rm m}&=&0.
\end{eqnarray}
The latter determines its evolution with the zero pressure $p_{\rm
m}=0$ as
\begin{equation}
\rho_{\rm m}=\frac{\rho_{\rm m0}}{a^3}.
\end{equation}
 Introducing the density parameters $\Omega_{\rm i}=\rho_{\rm
i}/3m^2_{\rm p}H^2$, which implies that the Friedmann equation
(\ref{fried}) can be rewritten as
\begin{equation} \label{omega}
\Omega_{\rm G}+\Omega_{\rm m}=1,
\end{equation} then one finds
\begin{equation}
\Omega_{\rm G}=\frac{n^2}{(H
\eta)^2}\Big[1+\Big(\frac{q^2}{n^2}\Big)\frac{1}{m^2_{\rm
p}\eta^2}\Big].
\end{equation}
The pressure is determined  by Eq.(\ref{cont1}) solely, using
$dx=Hdt$ as
\begin{equation}
p_{\rm G}=-\frac{1}{3}\frac{d\rho_{\rm G}}{d x}-\rho_{\rm
G}\end{equation}
 which  provides the EOS
 \begin{equation} \label{neos}
\omega_{\rm G}=\frac{p_{\rm G}}{\rho_{\rm G}}=-1+\frac{2 e^{-x}
\sqrt{\Omega_{\rm
G}}}{3n}\frac{\Big[1+2\Big(\frac{q^2}{n^2}\Big)\frac{1}{m^2_{\rm
p}\eta^2}\Big]}{\Big[1+\Big(\frac{q^2}{n^2}\Big)\frac{1}{m^2_{\rm
p}\eta^2}\Big]^{3/2}}.
 \end{equation}
In order to determine $\omega_{\rm G}$, we obtain the evolution
equation from the derivative of Eq.(\ref{omega}) with repect to
$t$ together with Eqs.(\ref{cont1}) and (\ref{cont2})
\begin{equation} \label{nequn}
\frac{d\Omega_{\rm G}}{dx}=-3\omega_{\rm G}\Omega_{\rm
G}(1-\Omega_{\rm G})
\end{equation}
and the relation of conformal time
\begin{equation} \label{etaeq}
\frac{d\eta}{dx}=\frac{1}{H_0}\Bigg[\frac{e^{-x}}{H/H_0}\Bigg]=
\frac{1}{H_0}\sqrt{\frac{e^x(1-\Omega_{\rm G})}{\Omega_{\rm m0}}}.
\end{equation}
Here we introduce the present Hubble parameter $H_0$ for our
purpose. For numerical computation, we rewrite Eqs.(\ref{neos})
and (\ref{etaeq})  by introducing $\zeta=H_0\eta$ as
\begin{equation} \label{eosze}
\omega_{\rm G}=-1+\frac{2 e^{-x} \sqrt{\Omega_{\rm
G}}}{3n}\frac{\Big[1+2\Big(\frac{H_0}{m_{\rm
p}}\Big)^2\Big(\frac{q^2}{n^2}\Big)\frac{1}{\zeta^2}\Big]}{\Big[1+\Big(\frac{H_0}{m_{\rm
p}}\Big)^2\Big(\frac{q^2}{n^2}\Big)\frac{1}{\zeta^2}\Big]^{3/2}}
\end{equation}
and
\begin{equation} \label{etazet}
\frac{d\zeta}{dx}=\sqrt{\frac{e^{x}(1-\Omega_{\rm G})}{\Omega_{\rm
m0}}}.
\end{equation}
The EOS of Eq.(\ref{eosze}) could be approximated as
\begin{equation} \label{neosa1}
\omega_{\rm G}\simeq -1+\frac{2 e^{-x} \sqrt{\Omega_{\rm
G}}}{3n}\Big[1+\frac{1}{2}\Big(\frac{H_0}{m_{\rm
p}}\Big)^2\Big(\frac{q^2}{n^2}\Big)\frac{1}{\zeta^2}\Big]
\end{equation}
for $(H_0/m_p)^2 \ll 1$. Actually, one has $(H_0/m_p)^2=3.03\times
10^{-122}h^2$ with $h \simeq 0.74$. This is just the ratio of the
energy density at present and Planck time $\rho_{0}/\rho_{\rm
p}=(l_{\rm p}/l_{\rm \Lambda})^2=(H_0/m_p)^2$, which reminds us
that the cosmological constant problem arises  if one introduces
the cosmological constant $\Lambda$ which satisfies $\omega_{\rm
\Lambda}=-1={\rm const}$~\cite{PR}. That is, observations needs to
have $ (l_{\rm p}/l_{\rm \Lambda})^2 \le 10^{-122}$, requiring
enormous fine-tuning of the cosmological constant from
1~\cite{Pad1,Pad2}. In this work, this fine-tuning is natually
included as a correction of the GUP in the EOS. This is because we
use a dynamical cosmological constant model of new agegraphic dark
energy model with the GUP.

Hence we expect that for the present and future dark
energy-dominated universe, the EOS is reduced to that of new
agegarphic dark energy model as~\cite{WC3,WC4,Neupane07,KLM}
\begin{equation} \label{neosa}
\omega_{\rm G}\to -1+\frac{2 e^{-x} \sqrt{\Omega_{\rm
G}}}{3n}=w_{\rm n}.
\end{equation}
Considering  Eq.(\ref{neosa})  together with the condition $a \to
0~(x \to-\infty)$ of the far past, the matter-dominated universe
is recovered with $\omega_{\rm n}=-2/3$ and $\Omega_{\rm
n}=n^2a^2/4$, while the radiation-dominated universe is recovered
with $\omega_{\rm n}=-1/3$ and $\Omega_{\rm n}=n^2a^2$~\cite{WC3}.
However, this prediction comes from the EOS $\omega_{\rm n}$ only.
We remind the reader that the pictures of far past and far future
should be determined from Eq. (\ref{nequn}) which governs the
whole evolution of the new agegraphic dark energy model~\cite{KLM}

 In the case of new
agegraphic dark energy model with the GUP, one finds from Fig. 1
that the whole evolution depends on parameter $n$ critically for
$q=1$. Here the initial condition is given by $\Omega_{\rm
G0}=0.72$ and $\eta_0=1/H_0(\zeta_0=1)$  at the present universe.
If $n$ is less than the critical value $n_c \sim 2.799531478$,
then its far past behavior is not acceptable because of $w_{\rm G}
\rightarrow \infty$ and $\Omega_{\rm G} \rightarrow 1$ as $ x
\rightarrow -\infty$. On the other hand, if $n$ is greater than
the critical value, then $w_{\rm G} \rightarrow -1$ and
$\Omega_{\rm G} \rightarrow 0$ as  $ x \rightarrow -\infty$. In
this case, we expect to have $\Omega_{\rm G} \propto a^2=e^{2x}$
from $w_{\rm G}$ in Eq.(\ref{eosze}). If $n$ approaches the
critical value, then $w_{\rm G} \rightarrow -2/3$ and $\Omega_{\rm
G} \rightarrow 0$ as $ x \rightarrow -\infty$. This corresponds to
the matter-dominated universe in the far past, predicted  by Wei
and Cai~\cite{WC3}. However, in the far future we have the
convergent results of $\Omega_{\rm G} \rightarrow 1,~\omega_{\rm
G} \rightarrow -1$, irrespective of $n$. This behavior is the same
as the new agegraphic dark energy models did show~\cite{KLM}.

\begin{figure}[t!]
\centering
\includegraphics{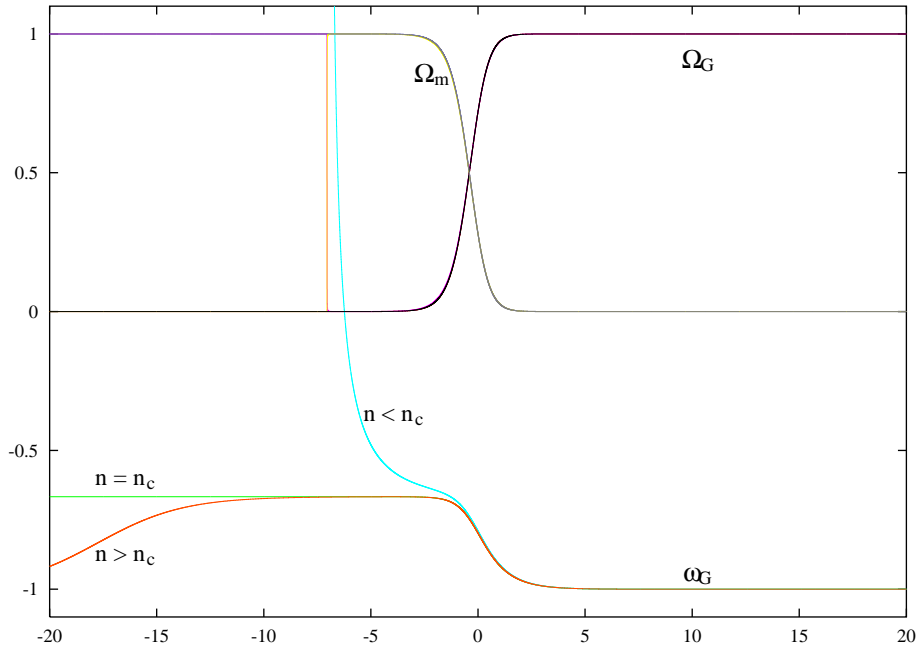}
\caption{ Graphs for the evolution of the new agegraphic dark
energy based on the GUP for $q=1$: $x$ vs $\Omega_{\rm
i},~\omega_{\rm G}$. From top to down, the density parameters
$\Omega_{\rm G}(\Omega_{\rm m})$ and the EOS $\omega_{\rm G}$ are
depicted  for $n < n_c~(\omega_{\rm G}\to \infty),
n=n_c=2.799531478 ~(\omega_{\rm G}\to -2/3)$ and $n
> n_c~(\omega_{\rm G}\to -1)$, respectively. } \label{fig1}
\end{figure}

\section{Discussions}
We discuss the effects of the GUP on the new agegraphic dark
energy models. The GUP is relevant to the Planck time, $t=t_{\rm
p}=10^{-43}s$. The GUP explains the cosmological constant problem
very well because it implies the Planck scale, $l=l_{\rm p}$.
Actually, the GUP does not change the present and future dark
energy-dominated universe significantly. In order to get the
Planck time behavior, the simulation must be being performed to
arrive $x=-120 $ from $x=0$. However, this task is formidable to
us and thus  we did not see what happens in this limit. Our
simulation was performed for the finite range of
$x\subset[-20,20]$ only.

In conclusion,  the new agegraphic dark energy model with the GUP
induces the Planck scale in the evolution of universe. However,
the GUP does not modify the present dark-energy dominated universe
significantly.

\section*{Acknowledgment}
H. Lee was  in part supported by KOSEF, Astrophysical Research
Center for the Structure and Evolution of the Cosmos at Sejong
University. Y. Myung  was in part supported by  the Korea Research
Foundation (KRF-2006-311-C00249) funded by the Korea Government
(MOEHRD).

        \end{document}